\documentclass[review]{elsarticle}

\usepackage{hyperref}
\usepackage{mathtools}
\usepackage{amssymb} 
\usepackage{graphicx}
\usepackage{subcaption}
\usepackage{CJKutf8}
\usepackage[bottom]{footmisc}
\usepackage{siunitx}

\usepackage[normalem]{ulem}
\useunder{\uline}{\ul}{}

\DeclarePairedDelimiter\set\{\}

\usepackage[table]{xcolor}
\definecolor{DeepGreen}{HTML}{6aa84f}

\journal{The Japanese Journal of the Institute of Industrial Applications Engineers}

\begin{document}

\begin{frontmatter}

\title{Relation Analysis between Hotel Review Rating Scores and Sentiment Analysis of Reviews by Chinese Tourists Visiting Japan}

\author[gidai]{Elisa Claire Alem\'an Carre\'on
\corref{mycorrespondingauthor}}
\ead{s153400@stn.nagaokaut.ac.jp}

\author[gidai]{Hirofumi Nonaka}
\ead{nonaka@kjs.nagaokaut.ac.jp}

\author[nagasaki]{Toru Hiraoka}
\ead{hiraoka@sun.ac.jp}

\address[gidai]{Nagaoka University of Technology, Nagaoka, Japan}
\address[nagasaki]{University of Nagasaki, Nagasaki, Japan}

\cortext[mycorrespondingauthor]{Corresponding author}

\begin{abstract}

In current times, the importance of online hotel review sites has become more and more apparent. Users of these sites reference of reviews strongly influences their purchase behavior and as such, reviews are important to companies and researchers alike. The majority of review sites offer both text reviews and numerical hotel ratings, and both information sources are widely used by researchers as a representation of a customer's sentiment and opinion. However, an opinion is a difficult concept to measure, and as such, depending on the relation these two sources have, it would be apparent whether or not it is safe to consider them equally in research. In this study we utilize an entropy-based Support Vector Machine to classify positive and negative sentiments in hotel reviews from the site \textit{Ctrip}, then calculating the ratio of positive and negative sentiment in each review and examine their correlation with said review's rating score using Spearman and Kendall Correlation coefficients and Maximal Information Coefficient (MIC).

\end{abstract}


\end{frontmatter}

\section{Introduction}\label{intro}

As the number of Chinese tourists visiting Japan increases, it is important for the hotel industry to conduct market research to analyze the needs of hotel guests. Under these circumstances, grasping needs through questionnaires and interviews has become the center of market research. However, surveys using such questionnaires and interviews have problems in terms of cost and real-time performance. On the other hand, with the spread of the Internet, there are many online reviews. Users will use the opinions of others as reference, and because of the large influence these have on purchase decisions \cite[][]{VERMEULEN2009123, SPARKS20111310}, these have also been actively used in the industry. Users' evaluations on many online review sites are numerous, and are divided into comment text as text information and a review score expressed numerically. The review scores are structured numerical information and are easy to use for analysis, so they are being used as evaluation indexes for users' products and services \cite[][]{XIE20141, BULCHANDGIDUMAL201344, ZHOU20141}. On the other hand, text reviews are also being analyzed based on natural language processing. For example, research that analyzes sentiment of word-of-mouth using information theory \cite[][]{AMPLAYO201754} as well as forecast of product sales \cite[][]{FAN201790} or the ranking of products based on sentiment analysisis \cite[][]{LIU2017149} is being conducted.

In customer trend analysis, what is used as an analysis index is extremely important. As mentioned above, in previous studies, customer behavior was analyzed using sentiment analysis only \cite[][]{AMPLAYO201754,LIU2017149} and there are many studies that use only the review score points \cite[][]{XIE20141, BULCHANDGIDUMAL201344, ZHOU20141}. Therefore, it is necessary to examine the relationship between these review scores and the sentiment analysis of comments. If the relationship between the review score and the sentiment analysis of the document is high, it would be easy and there would be a great merits to do analysis based only on the numerical information or using the numerical information as training data for future sentiment analysis. On the other hand, if the correlation is low, it is necessary to comprehensively evaluate the reviews by using the score and text information together. Alternatively, it is necessary to select a method that more reflects the user's emotions and opinions and use it as an evaluation index. In this way, investigating the relationship between the review score and the sentiment analysis of the comment text is extremely important in the analysis of the review.

Therefore, in this study, we investigated the relationship between the review scores and the sentiment analysis of the comment text. In this study, we first collected a large number of review documents written about Japanese hotels and their review score from the online hotel review site \textit{Ctrip} for Chinese tourists. Next, we trained an SVM using the entropy-based feature selection method developed by our researchers, and classified documents that express positive emotions and documents that express negative emotions. Since the feature vector characteristic of this emotion classification was constructed by keyword extraction based on entropy, the classification was performed using a language-independent method based on statistics. Furthermore, based on the emotion classification of each sentence in each review, the ratio of sentences expressing satisfaction and the ratio of dissatisfaction sentences to the full review were calculated in order to quantify emotions. Finally, the interrelationship between the review score and the emotional evaluation of the review was analyzed from Spearman's rank correlation coefficient and Kendall's rank correlation coefficient MIC. This will be described in detail below.

\section{Previous Work}\label{lit_rev}

In previous research, targeting product reviews is the mainstream. For example, a study that applied Word Cloud to extract words that are often used by consumers \cite[][]{hargreaves2015}, and a market analysis using sentiment analysis of product reviews using an emotion dictionary called HowNet \cite[][]{zhang2011feature}. On the other hand, as for quantitative evaluation of the impact of hotel online reviews, there is a study that proves that hotel online reviews have a particular effect on customer motivation \cite[][]{VERMEULEN2009123}. There are other studies that have shown a relationship between sales and the textual information of reviews that are evaluated \cite[][]{basuroy2003}. In the mentioned studies, the scores were not considered, and the relationship between the scores and the text information was not focused on.

\section{Methodology}\label{methodology}

\subsection{Preprocessing}\label{preprocessing}

At the crawling stage, we used the fact that the URL structure is determined by the hotel ID number, and so each hotel page can be automatically loaded. Next, scraping was performed, and the documents, IDs, review score, etc. of each review were acquired using the structure of the HTML code and saved in our database. Morphological analysis was also performed for the review. As a tool for morphological analysis of Chinese, Stanford Word Segmenter \cite[][]{chang2008} provided by The Stanford NLP Group of Stanford University was used.

\subsection{Sentiment Analysis}\label{sentiment_analysis}

Samples were extracted from the data collected by online hotel reviews for sentiment analysis, and with the cooperation of three Chinese research students, each sentence in each review was tagged according to whether it expressed satisfaction or dissatisfaction. Manual classification of Positive and Negative was performed, and training data was created. Then, the words and emotion classification tags included in the sample review were trained by SVM, and the emotion classification of all the data in the population was performed. A feature vector characteristic of emotion classification by SVM was constructed by keyword extraction based on entropy, which will be described later. The details will be described below.

\subsubsection{Entropy Based Keyword Extraction}\label{entropy}

\begin{figure}[bh]
    \centering
    \begin{subfigure}[b]{0.4\linewidth}
        \includegraphics[width=\linewidth]{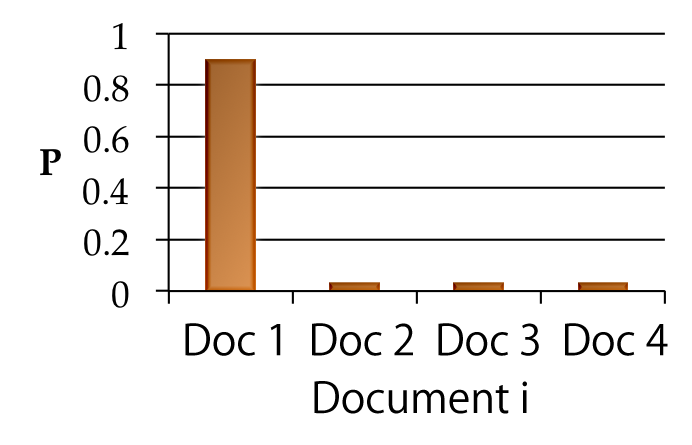}
        \caption{Entropy close to zero.}
    \end{subfigure}
    \begin{subfigure}[b]{0.4\linewidth}
        \includegraphics[width=\linewidth]{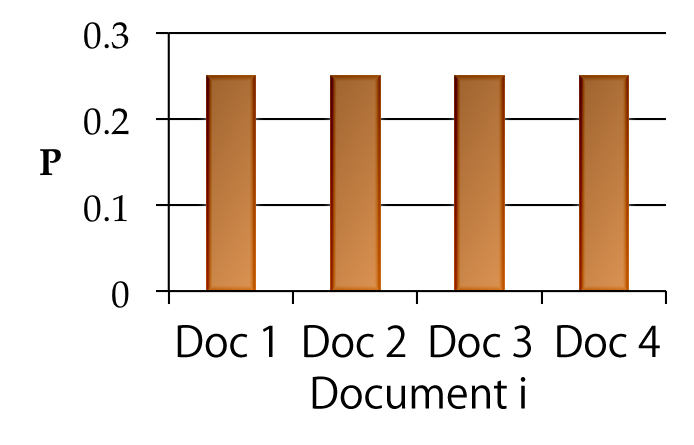}
        \caption{High entropy.}
    \end{subfigure}
\caption{Probabilities of a word \(j\) being contained in a document \(i\).}
\label{fig:entropygraphs}
\end{figure}

In this study, we based the extraction of the keywords that are influenced by the users’ emotional judgement on the calculation of an entropy value for each word. Speaking in Information Theory terms, Shannon’s Entropy is the expected value of the information content in a signal \cite[][]{shannon1948}. Applying this knowledge to the study of words allows us to observe the probability distribution of any given word inside the corpus. For example, a word that keeps reappearing in many different documents will have a high entropy, given that predicting on which document it would appear becomes uncertain. On the contrary, a word that only was used in a single text and not in any other documents in the corpus will be perfectly predictable to only appear in that single document, bearing an entropy of zero. This concept is shown in Fig.\ref{fig:entropygraphs}.

Based on the meaning of entropy explained above, keywords that will be considered positive will have a large entropy when they appear in many positive documents, and a smaller entropy in negative documents. The same will occur for negative keywords in the opposite documents. In this study we use the entropy values of keywords to perform a classification. First, we tagged a set of documents as positive or negative. Then, for each word \(j\) that appears in each document \(i\), we counted the number of times a word appears in positive comments as \(N_{ijP}\), and the number of times a word appears in negative comments as \(N_{ijN}\). Then, as shown in the formulas below, we calculated the probability of each word appearing in each document shown below as \(P_{ijP}\) (\ref{eq:PijP}) and \(P_{ijN}\) (\ref{eq:PijN}).

\begin{equation}\label{eq:PijP}
P_{ijP} = \frac{N_{ijP}}{\sum_{i=1}^M N_{ijP}}
\end{equation}

\begin{equation}\label{eq:PijN}
P_{ijN} = \frac{N_{ijN}}{\sum_{i=1}^M N_{ijN}}
\end{equation}

We then substitute these values in the next formula. We calculated the entropy for each word \(j\) in relation to positive documents as \(H_{Pj}\) (\ref{eq:Hpj}), and the entropy for each word \(j\) in relation to negative texts as \(H_{Nj}\) (\ref{eq:Hnj}). That is, as is shown in (\ref{eq:lim_Hpj}) and (\ref{eq:lim_Hnj}), all instances of the summation when the probabilities \(P_{ijP}\) or \(P_{ijN}\) are zero and the logarithm of these becomes undefined are substituted as zero into (\ref{eq:Hpj}) and (\ref{eq:Hnj}).

\begin{equation}\label{eq:Hpj}
H_{Pj} = - \sum_{i=1}^M [P_{ijP}\log_2 P_{ijP}]
\end{equation}

\begin{equation}\label{eq:lim_Hpj}
\lim_{P_{ijP}\to0+} P_{ijP}\log_2 P_{ijP} = 0
\end{equation}

\begin{equation}\label{eq:Hnj}
H_{Nj} = - \sum_{i=1}^M [P_{ijN}\log_2 P_{ijN}]
\end{equation}

\begin{equation}\label{eq:lim_Hnj}
\lim_{P_{ijN}\to0+} P_{ijN}\log_2 P_{ijN} = 0
\end{equation}

After calculating the entropies for each word, we adjusted for their \(\alpha\) value by testing for the highest F-value. A positive keyword is determined when (\ref{eq:entropy_pos}) is true, and likewise, a negative keyword is determined when (\ref{eq:entropy_neg}) is true for the best performing \(\alpha\) value, using these keywords as elements for training an SVM \cite[][]{cortes1995}. The performance was determined using a k-fold cross validation calculating the best \(F_1\) value \cite[][]{powers2011}.

\begin{equation}\label{eq:entropy_pos}
H_{Pj} > \alpha H_{Nj}
\end{equation}

\begin{equation}\label{eq:entropy_neg}
H_{Nj} > \alpha' H_{Pj}
\end{equation}

\subsection{Correlation Analysis}\label{correlation}

The following method was experimentally used to measure the correlation of the ratio of positive sentiment obtained by dividing the sentences judged as positive included in each review \(x\) to the review scores \(y\).

\subsubsection{Pearson correlation coefficient \(r\)}\label{correl_pearson}

In order to measure the correlation of the sentiment ratio \(x\), obtained by dividing the number of sentences judged as positive included in each review \(i_P\) by the total number of sentences \(i_T\), to the score \(y\), one of the methods used was Pearson's correlation coefficient \(r\) (\ref{eq:pearson_r}). The formula is shown below. The value of the formula (\ref{eq:pos_neg_ratio}) is substituted into the formula (\ref{eq:pearson_r}). When calculating the negative rate, substitute the number of sentences judged to be negative \(i_N\) into the formula (\ref{eq:pos_neg_ratio}).

\begin{equation}\label{eq:pos_neg_ratio}
x = \frac{i_P}{i_T}
\end{equation}

\begin{equation}\label{eq:pearson_r}
r = \frac{{}\sum_{i=1}^{M} (x_i - \overline{x})(y_i - \overline{y})}
{\sqrt{\sum_{i=1}^{M} (x_i - \overline{x})^2(y_i - \overline{y})^2}}
\end{equation}

\subsubsection{Spearman's rank correlation coefficient \(\rho\)}\label{correl_spearman}

In order to measure the correlation of the sentiment ratio \(x\), obtained by dividing the number of sentences judged as positive included in each review \(i_P\) by the total number of sentences \(i_T\), to the score \(y\), since twe consider the score to be a ranked variable, we used Spearman's ranked correlation coefficient \(\rho\) (\ref{eq:spearman_rho}) which is also based on Pearson's correlation coefficient. The formula is shown below. Substitute the value of the formula (\ref{eq:pos_neg_ratio}) into the formula (\ref{eq:spearman_rho}).

\begin{equation}\label{eq:spearman_rho}
r_s = \rho_{rg_X,rg_Y} = \frac{cov(rg_X,rg_Y)}{\sigma_{rg_X} \sigma_{rg_Y}}
\end{equation}

\subsubsection{Kendal's rank correlation coefficient \(\tau\)}\label{correl_kendall}

Like Spearman's rank correlation coefficient, Kendal's rank correlation coefficient is used to investigate the relationship between the values that represent rank. The formula (\ref{eq:kendall_tau}) is shown below. However, substitute the expression (\ref{eq:kendall_L}) and the expression (\ref{eq:kendall_K}) into the expression (\ref{eq:kendall_tau}). Substitute the expression (\ref{eq:pos_neg_ratio}) for each.

\begin{equation}\label{eq:kendall_tau}
\tau = (K-L)/\dbinom{n}{2}
\end{equation}

\begin{equation}\label{eq:kendall_L}
L = \# \set[\Big]{\{i,j\} \in \binom{[n]}{2} \mid \neg (x_i \lessgtr x_j, y_i \lessgtr y_j ) }
\end{equation}

\begin{equation}\label{eq:kendall_K}
K = \# \set[\Big]{\{i,j\} \in \binom{[n]}{2} \mid (x_i \lessgtr x_j, y_i \lessgtr y_j ) }
\end{equation}

\subsubsection{MIC}\label{mic}

\begin{figure}[bh]
\centering
\includegraphics[width=0.9\linewidth]{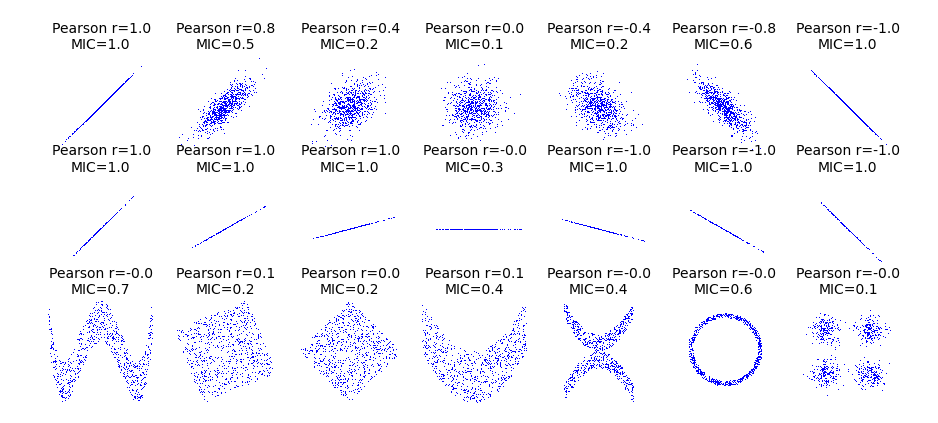}
\caption{Comparison of minepy MIC and Pearson \(r\) in various cases}
\label{fig:mic}
\end{figure}

Pearson's correlation coefficient can only extract linear relationships. On the other hand, there is MIC (Maximal Information Coefficient) as a method to analyze the relationship between two variables including non-linearity \cite[][]{Reshef2011}. It is an index for analyzing the relationship between variables including non-linearity based on the amount of mutual information, considering the two variables for which you want to analyze the relationship as random variables. Fig. \ref{fig:mic} shows several examples of comparing the coefficients of MIC and Pearson. Pearson's coefficient is a value from 0 to 1 in a linear relationship, and it is determined whether it is a positive value or a negative value depending on the direction of inclination. On the other hand, in the case of MIC, even if it is a non-linear relationship, it can express the correlation using the value from 0 to 1 as long as there is a relationship. Examples of this are shown in Fig.\ref{fig:mic}.

The procedure for calculating MIC will be described. First, for the variables \(X\) and \(Y\) to be analyzed, after plotting the two variables on the coordinate space, the space is divided by \(a * b\) (Split the \(X\) direction into \(a\) parts, and the \(Y\) direction into \(b\) parts). Then, for each of the two variables, the cell existence probability can be calculated by dividing the number of sample points belonging to each cell by the total number of samples. That is, \(X\) and \(Y\) are regarded as random variables based on the existence probability in the cell. This makes it possible to calculate the Mutual Information for \(X\) and \(Y\).

At this time, even if the original two variables have a non-linear relationship as well as a linear relationship, the dependency between the random variables becomes strong, so the mutual information takes a large value. Therefore, unlike Pearson's correlation coefficient, it is possible to extract a non-linear relationship. Now, with the MIC, the width and length of each cell are unequally spaced, so there are innumerable division methods (note that each cell has a maximum resolution). For the purpose of extracting non-linearity, it is necessary to find a dividing grid that maximizes the amount of mutual information as much as possible. In MIC, it is assumed that \(X\) is now divided into arbitrary \(a\) pieces and \(Y\) is divided into arbitrary \(b\) pieces. At this time, we find a division that maximizes the amount of mutual information in the division of \(a * b\) by brute force. The formula (\ref{eq:mic}) for mutual information is shown below.

\begin{equation}\label{eq:mic}
I(X;Y) = \int_{Y}{\int_{X}{p(x,y)\log(\frac{p(x,y)}{p(x)p(y)})}\,dx}\,dy
\end{equation}

In this study, the MIC was calculated using the Python library minepy \cite[][]{Albanese2012}. The Fig. \ref{fig:mic}, which compares Pearson's correlation coefficient and MIC, was obtained from the minepy API site.

\section{Experiment Results}\label{results}

The following describes what was specifically done in this study using the method described earlier.

\subsection{Preprocessing}\label{res_preprocessing}

First, we collected 1,541,424 HTML files from May 2016 to September 2016 from \textit{Ctrip}. Of the 1,541,424 HTML files, we collected data on 5,938 hotels in Japan. Among them, he scraped the comment text of 44,912 reviews. 286,109 sentences divided into sentence units are analyzed. In addition, the scores of the reviews were also collected.

\subsection{Sentiment analysis performance}\label{res_sentiment_analysis}

In order to create training data, a sample of reviews was randomly extracted from all the data, divided into sentences by expressing satisfaction or dissatisfaction, tagging work of 159 sentences was performed manually. The entropy was calculated from the sentences that become the training data.

After calculating the entropy for the training data, the features with the maximum \(F_1\) value were selected by evaluating \(\alpha\) from 1.0 to 3.75 with a step size of 0.25 in order to obtain its optimum value. After training the SVM, the keyword list was selected based on the \(\alpha\) which led to the maximum \(F_1\) value as a result of 5-Fold Cross Validation (\(k = 5\)) for the evaluation data. Furthermore, both lists were combined to create a new list, and 5-Fold Cross Validation was performed in the same manner. The evaluation results are shown in Table \ref{tab:kfold_performance}. The feature that combines both has the highest \(F_1\) value, and \(F_1\) classifies all data with an SVM with a high accuracy of 0.95. Positive keywords are, for example, "\begin{CJK}{UTF8}{min}热情\end{CJK}" and "\begin{CJK}{UTF8}{min}景色\end{CJK}" (indicating "friendly" and "(good) scenery" respectively), and in the case of Negative keywords, there was an example of "\begin{CJK}{UTF8}{min}价格\end{CJK}" or "price" (indicating dissatisfaction because it is high).

\begin{table}[th]
\centering
\caption{5-fold Cross Validation performance results}
\label{tab:kfold_performance}
\begin{tabular}{llll}
Keyword list                & C            & \(F_1 \mu\)       & \(F_1 \sigma\) \\
Positive keywords (\(\alpha=2.75\))  & 2.5          & 0.91                & 0.01           \\
Negative keywords (\(\alpha'=3.75\)) & 0.5          & 0.67                & 0.11           \\
\rowcolor{DeepGreen}\color{white}\textbf{Combination}
                                     & \textbf{0.5} & {\ul \textbf{0.95}} & \textbf{0.01}
\end{tabular}
\end{table}

\begin{table}[th]
\centering
\caption{Correlation of sentiment analysis and score results}
\label{tab:res_correl}
\begin{tabular}{llll}
\textbf{Sentence ratio} & \textbf{Spearman's \(\rho\)} & \textbf{Kendall's \(\tau\)} & \textbf{MIC} \\
Positive ratio     & 0.161            & 0.125            & 0.049        \\
Negative ratio     & -0.149           & -0.122           & 0.0447      
\end{tabular}
\end{table}

After learning with the optimal model, sentiment analysis was performed on unknown data, and the above-mentioned "Positive ratio" and "Negative ratio" were calculated. Since these values are the ratio of sentences expressing satisfaction and sentences expressing dissatisfaction, they were treated as numerical coefficients expressing emotions in each sentence, and correlation analysis was performed.

\subsection{Correlation analysis}\label{res_correls}

The positive and negative rates of each review in all the data and the review score of the reviews were analyzed using Spearman's rank correlation coefficient, Kendall's rank correlation coefficient, and MIC. The results are shown in Table \ref{tab:res_correl}.

\section{Discussion}\label{discussion}

It was shown that both the Positive ratios and the Negative ratios were very low in relation to the review score in all the indicators.

Therefore, since the relationship between the result of sentiment analysis in the comment text and the review score is low, when analyzing the user's opinion from the review, considering both the content of the text and the numerical review score and their differences is very important.

In the past, although this relationship has not been shown, only review scores are often used as indicators of satisfaction and emotional evaluation. For example, the studies by Xie et al. And Burchand-Gidumal et al. used the scores for the hotel as a proxy for satisfaction\cite[][]{XIE20141, BULCHANDGIDUMAL201344}, and Zhou et al. investigated the factors of satisfaction using a multivariate analysis to do this, but the dependent variable was the review score \cite[][]{ZHOU20141}.
Based on the results of this study, it was suggested that it is not appropriate to use only the review score or sentiment analysis as an index to measure the satisfaction of tourists.

\section{Conclusion}\label{conclusion}

In this study, we investigated the relationship between the results of sentiment analysis in the text text of online hotel reviews and the evaluation points. We constructed a method with high classification performance (\(F_1 = 0.95\)) for sentiment analysis, and calculated the ratio of sentences classified as Positive and the ratio of sentences classified as Negative for each review. For the relationship, Spearman's rank correlation coefficient, Kendall's rank correlation coefficient, and MIC were used. As a result, all showed low values. Therefore, it was clarified that the relationship between the result of sentiment analysis in the comment text and the evaluation point, which is numerical information, is low. Therefore, it was considered that a more comprehensive evaluation was important. In the future, based on these results, we will investigate whether the result of sentiment analysis or the evaluation score expresses the user's opinion more after comparing with the analysis using multilingual information, and comprehensively reviewing. We will proceed with the development of analysis methods.

\section*{Acknowledgements}\label{acknowledgements}

We would like to thank Mr. Liangyuan Zhou and Ms. Eerdengqiqige for their support in creating teacher data in Chinese. In addition, this paper was supported by the "Japan Construction Information Center".

\clearpage

\bibliography{jjiiae-scores_en}

\end{document}